# Metal-Graphene Heterojunction Modulation via $H_2$ interaction

A. R. Cadore,[1,a] E. Mania,[1] E.A. Morais,[2] K. Watanabe,[3] T. Taniguchi,[3] R. G. Lacerda[1] and L. C. Campos[1,a]

[1]*Departamento de Física, Universidade Federal de Minas Gerais, Belo Horizonte, 30123-970, Brazil*

[2]*Universidade Federal de Itajubá, Campus Avançado de Itabira, Itabira, 35903-087, Brazil*

[3]*Advanced Materials Laboratory, National Institute for Materials Science, 1-1Namiki, 305-0044, Japan*

[a] *Electronic mail: alissoncadore@gmail.com; lccampos@fisica.ufmg.br*

Combining experiment and theory, we investigate how a naturally created heterojunction (*pn* junction) at a graphene and metallic contact interface is modulated via interaction with molecular hydrogen ($H_2$). Due to an electrostatic interaction, metallic electrodes induce *pn* junctions in graphene, leading to an asymmetrical resistance for electronic transport via electrons and holes. We report that an asymmetry in the resistance can be tuned in a reversible manner by exposing graphene devices to $H_2$. The interaction between the $H_2$ and graphene occurs solely at the graphene-contact pn junction and might be due to a modification on the electrostatic interaction between graphene and metallic contacts. We confront the experimental data with theory providing information concerning the length of the heterojunction, and how it changes as a function of $H_2$ adsorption. Our results are valuable for understanding the nature of the metal-graphene interfaces and point out to a novel route towards selective hydrogen sensor application.

Graphene is a zero-gap semiconductor which charge carrier density and conductance can be controlled electrostatically by preparing graphene devices as field effect transistors.[1,2] In such architecture, the contact resistance considerably impairs device performance and is responsible for a conduction asymmetry for *p*-doped and *n*-doped graphene.[3–8] This asymmetry stems from heterojunctions (*pn* junctions) formed at metal-graphene interfaces due to different work functions between graphene and metal – i.e. Fermi level pinning.[9–13] Effectively, the advent of the *pn* junction results in an additional charge scattering at the metal-graphene interface increasing device resistance. On the other hand, interesting effects can be observed as well. For instance: heterojunctions at metal-graphene interface have been used to design Fabry-Perot cavities[11] and to observe resonances in Josephson junctions in graphene devices.[14] In both systems, the metal-graphene interfaces have been considered a static problem, where the doping at the graphene underneath the contact is solely defined by the type of metal used. However, so far, a controllable method to probe and modulate the *pn* junction induced by contacts has not been reported. In addition, there are still open questions about how far a *pn* junction can extend from the contact into the graphene channel and if there are technological applications based on specialties of graphene contact resistance.[3,4,6,7,14–16]

In this work, we show that molecular hydrogen (H$_2$) can reversibly modulate the heterojunction at the metal-graphene interface. Our experimental and theoretical approaches demonstrate that the heterojunction extends over few nanometers and that the electrostatic modulation of the interface is uniquely possible for small and polarizable molecules such as H$_2$. Moreover, this discovery is valuable for understanding the metal-graphene interface and opens up applications for selective detection of H$_2$ without chemical or physical functionalization.[17]

Contact resistance in graphene devices has being extensively investigated.[3–5,10,18,19] It is known that graphene devices produced in a Hall bar geometry (non-invasive contacts) show a more symmetrical conductance dependence with the gate voltage in both electron and hole branches.[5] Conversely, in two-terminal, or in a four-probe geometry with contacts that go across the entire cross section of the device (invasive contacts), electron and hole conductance is essentially asymmetrical, which occurs as a consequence of the formation of a *pn* junction at the metal-graphene interface.[4,5,9,11,18] In order to investigate the conductance modulation in graphene devices, we perform experiments on high quality graphene/*h*-BN devices showing charge mobility of the order of $\mu \sim 15{,}000$ cm$^2$/Vs at room temperature.

We prepare graphene from the standard scotch tape method. Few-layer graphite is exfoliated on SiO$_2$ (285nm thick) to be used as a back gate bias. Then, we transfer *h*-BN flakes first,[20] followed by graphene atop of it, forming a graphene/*h*-BN/graphite heterostructure. We also study graphene devices on top of *h*-BN/SiO$_2$ substrate, where doped Si substrate acts as a back gate. In these devices, resistance is modulated by H$_2$, as well as in graphene/*h*-BN/graphite devices, however, at temperatures higher than 100 °C we observe charging effects caused by possible trapped charges at *h*-BN/SiO$_2$ interface.[21] Therefore, all data shown here are taken from graphene/*h*-BN/graphite architecture. The flatness and cleanliness of the graphite, graphene and *h*-BN are verified by atomic force microscope (AFM), and identification of graphene samples is made by optical analysis and by Raman spectroscope. After each material transfer, samples are submitted to an standard heat cleaning process in Ar/H$_2$ at 350°C to remove organic residues.[20] Electron-beam lithography and oxygen plasma etching are used to define the graphene geometry, and thermal evaporation of Cr/Au is used to fabricate the contacts (1/50nm). Finally, to remove polymer residues reminiscent from the lithography processes, devices are submitted to another heat cleaning process.

Since our measurements are not performed in vacuum, an initial standard procedure is taken to remove humidity from the samples.[22] Afterwards, the charge neutrality point (CNP) of our devices stays close to zero back gate voltage, indicating minor doping. Voltage measurements between probes are performed using standard lock-in techniques at a frequency of 17Hz, with a current bias of 100nA applied between source and drain, as it is indicated in Fig. 1. All measurements shown in this paper are performed at $T=230$°C – temperature at which the changes on resistance due to hydrogen interaction are more pronounced (see Supp. Mat.). As we show in Fig. 1(a), the resistance of a graphene device with non-invasive contacts tends to be symmetrical between the hole-branch ($V_\text{G} < V_\text{G}^\text{D}$) and electron-branch ($V_\text{G} > V_\text{G}^\text{D}$); $V_\text{G}$ is the applied back gate voltage, and $V_G^D$ is the point that shows the resistance maxima. Meanwhile, if contacts are placed in the configuration shown at Fig. 1(b) (red curve), device resistance is asymmetrical, showing larger resistance at the electron-branch when compared with the hole-branch for the same density of charge. In this invasive geometry, current has to flow through the metal-graphene interface and scattering due to the *pn* junction cannot be ignored. In general, electrical properties observed in graphene devices with invasive gold contacts indicate the presence of *pn* junctions at the contact region due to a *p*-type doping of graphene underneath the metal.[3,5,18]

We now perform the same experiment adding $H_2$ into the chamber. Firstly, we set initial conditions to the experiment applying a flow of ultra-pure Ar (500sccm), then we insert ultra-pure $H_2$, keeping the total gas flow constant at 500sccm. Fig. 1(a) shows total resistance ($R_T$) as a function of gate voltage for a non-invasive four-probe device. In red, we depict the total resistance under pure Ar atmosphere (this data is identical to experiments performed in vacuum), and in black (dashed line) we show data taken under Ar+$H_2$. For devices with non-invasive contacts, there is no significant change on the graphene resistance due to $H_2$ exposure. Indicating that the molecular hydrogen does not transfer charge or react with the graphene channel, as expected.[23,24] Meanwhile, for invasive configuration, the total resistance changes in an asymmetric manner when the sample is exposed to $H_2$. Fig. 1(b) depicts how the device resistance decreases at the electron-branch (*n*-type) and increases at the hole-branch (*p*-type) when a sample is exposed to a fixed concentration of 20% of $H_2$ (gradient from red to blue lines in the Fig. 1(b)). In this figure, $t_0$ depicts the data taken in pure Ar before $H_2$ exposure, and $t_F$ is the final data taken under $H_2$ exposure (~30min after inserting $H_2$). The reversibility shown in the inset of Fig. 1(b) suggests that neither hydrogenation nor formation of permanent bonding occur. Indeed, we confirm the graphene integrity in graphene/*h*-BN/$SiO_2$ samples by Raman spectroscopy after the electrical measurements, and no significant "defective" D peak is observed.

Our findings show that graphene devices with invasive contacts can be used to detect $H_2$ without any kind of graphene functionalization, but we leave a more precise sensor characterization for further works. However, it is worth mention some details about $H_2$ detection. Changes on contact resistance show to be thermally activated, exhibiting an enhancement at higher temperatures (see Supp. Mat.). In addition, the device response time – total amount of time necessary for the device to change from its previous state to a final state with a tolerance of 10% – is highly sensitive to hydrogen concentration. For instance, we perform experiments changing $H_2$ concentration and the response time is around 10min for 50% of hydrogen, and 40min for 0.1%. However, in order to avoid high concentration of $H_2$ in our chamber, we perform most of our experiment using 20% of $H_2$ as we present in this work.

Asymmetries between graphene device resistance at the electron and hole-branch can be assigned to various reasons like different electron (hole) charge scattering cross-section due to impurities, charge inhomogeneity or contact resistance effects.[3,9,14,18] However, our experiments confirm that, at high quality samples, the contact resistance is responsible for the resistance asymmetries and molecular hydrogen is able to tune the scattering rate at the metal-graphene interface. This tuning process caused by the $H_2$ molecules will be discussed in detail later. Let us first verify that Boltzmann theory captures the main features of the graphene/$h$-BN device at high temperatures. Fig. 1(a) shows that, as expected, devices with non-invasive leads show symmetrical resistance for both carriers, and contact resistance can be ignored. In addition, as we show in Fig. 2(a), Boltzmann theory is a good approach to describe the data.[25,26] Here, the main channel resistance is calculated by: $\sigma^{-1} = \rho_s + (ne\mu + \sigma_o)^{-1}$; $\mu$ is the maximum charge mobility; $\rho_s$ is the contribution to resistivity from short range scattering; and $\sigma_o$ is the residual conductivity at the CNP. The discrepancy between data and Boltzmann theory near of CNP is due to thermo-activated charges. However, Boltzmann theory converges to the experiments near of the CNP at low temperatures (see Supp. Mat.).

In the case of devices containing invasive contacts, there are significant contributions of the electrodes, i.e., heterojunctions at metal-graphene interfaces. Thus, to model the total device resistance, we have to consider electrostatic effects at the interface between graphene and contacts. In our approximation, we consider two contributions: the resistance of the main graphene channel and the contact resistance (more details see Supp. Mat.). The main channel is described by Boltzmann theory, as discussed above, while the contact resistance has contributions of the classical Ohm's law, as well as a quantum effects. This quantum contribution addresses scattering due to the *pn* junction at the interface ($R_{pn}$). Furthermore, for our four-probe terminal devices we do not consider the contribution of the classical contact resistance.[4]

$$R_T = R_{pn} + R[L_o - L] \qquad (1)$$

Here $R_T$ is the total resistance of the device; $R[L_o\text{-}L]$ is the main channel resistance calculated by Boltzmann theory; $L$ and $L_o$ are the lengths of the graphene channel and the heterojunction, respectively; $R_{pn}$ is the resistance ascribed for the *pn* junction. The resistance of the interface $R_{pn}$ is calculated via the Landauer formula $\left(R_{pn} = \left[\left(\frac{4e^2}{h}\right)\sum_{k_y} T(k_y)\right]^{-1}\right)$, where $e^2/h$ is the quantum of conductance and the sum incorporates the transmission probability $(T(k_y))$ of each propagating mode $(k_y)$ across the metal-graphene interface.[27–29]

In Fig. 2(a) and 2(b), we show the resistance across the graphene devices measured in atmosphere of argon for non-invasive and invasive contacts, respectively. With black circles, we show the fitting performed using our model described by equation 1. The good agreement between experiment and theory reinforces the existence of the *pn* junction at devices with invasive probes, while, for non-invasive electrodes, the influence of the *pn* junction is not observed. By performing a fitting of the theory on our experiments, we estimate the length of the *pn* junction interface. At ordinary devices, our best fitting indicates that the length of the interface is about $L_{max}$~10nm long at $n = 2\times10^{12}$cm$^{-2}$. However, for the region close to CNP it is not possible to estimate the interface length due to energy fluctuation associated to the temperature (~40 meV at $T$=230°C) and graphene disorder (~30meV estimated at $T$=4K). The interface length is in accordance with other works that indirectly estimate it.[3,11,14] Also, the mean free path ($l_{mpf} = \sigma h/2e^2 k_F$) determined from the conductivity ($\sigma$) at $n = 2 \times 10^{12} cm^{-2}$ is found to be about 60nm. Therefore, in graphene devices with interface length around 10nm, it is reasonable to consider ballistic transport across the heterojunction.[29]

Finally, to understand how $H_2$ modifies the length of this interface and the Fermi level pinning, we focus on the electrostatic interaction between graphene and contacts. It is expected that gold contacts stay at an equilibrium separation of about 0.3nm up to 0.5nm from the graphene layer.[13] Thus, Au is weakly adsorbed over graphene, preserving its band structure; but, due to work function differences, charges can be transferred to/from the graphene, shifting the Fermi energy.[4] The amount of charge transferred depends on the difference between the work functions of graphene and the metal, as well as on the equilibrium separation, defining the doping at the graphene underneath the contacts.[12,13] In addition, the imbalance between the charge density at the graphene underneath the metal and the graphene main channel results in an in-plane electrostatic profile in the graphene,[13,30] leading to a formation of a *pn* junction with a length *L* along the graphene main channel (see insets in Fig 2 (d)).

Now, we analyze the case where the metal-graphene interface is exposed to $H_2$ molecules. The size of a molecular hydrogen is expected to be about 0.15nm, which is smaller than the gold contact equilibrium separation.[23] So it is plausible to expect that some molecules are allowed to diffuse and position between the metal-graphene interface. This hypothesis is also likely to happen, since gold has a low $H_2$ diffusion coefficient[31] and does not react[32] with $H_2$. Also, our non-invasive experiment clearly shows that $H_2$ does not interact with the main graphene channel (Fig. 1(a)). This analysis supports that the conductance changes due to $H_2$ exposure only happen at the metal-graphene interface in the invasive configuration. Therefore, it is possible that the molecules change the interface potential due to the creation of a dipole layer at the interface between graphene and contact[31,33] or changing the equilibrium separation as well.[12,13] However, the simulation of this phenomena is out of the scope of this work.

In this context, our results help to better understand the *pn* junction at the metal-graphene interface. Besides that, it shows that the hydrogen considerably modifies the Fermi level pinning induced by gold contacts, causing an inversion in the doping – from *p*-type to *n*-type. The contact resistance modulation under $H_2$ exposure is shown in Fig 2. Fig. 2(c) shows the resistance asymmetry between the charge carriers. $R_{odd}$ is defined as the difference between the device resistance of the electron-branch and the hole-branch at the same charge density:[5] $R_{odd}(\Delta V_G) = \frac{1}{2}[R(V_G^D + \Delta V_G) - R(V_G^D - \Delta V_G)]$. Initially, we observe a higher device resistance for electrons compared with holes, which is consistent with an existence of a *pn* junction at the contact interface; graphene main channel is *n*-type doped and graphene near the contacts are *p*-type doped. Nevertheless, the inversion of $R_{odd}$ indicates that the metallic contact is now performing an *n*-type doping instead, and confirms that the presence of $H_2$ modifies the Fermi level pinning induced by the contact from *p*-type to *n*-type (see inset Fig. 2(d)). Besides that, a small shift of the CNP ($\Delta V = 0.18$ V) can be noticed in Fig. 1(b). This shift can not be addressed to charge transferring from $H_2$ to graphene, since it is only observed at devices with invasive electrodes. Therefore, we believe that the shift of the CNP is caused by a change on the average density of charge due to the inversion of the doping underneath the contacts.[12,13]

Another property of the heterojunction which we tried to roughly estimate is the length of the *pn* junction. The length (*L*) depends mainly of the unknown doping level underneath the contact and the doping at the main graphene channel, as it is illustrated by the inset in Fig. 2(d). We estimate $L$~10nm at $V_G$=2V assuming a symmetrical heterojunction (*n*-type doping in the main channel is equal to *p*-type doping underneath the contacts). This value is in agreement with previous works,[3,11,14] and it gives us a rough idea of how much the heterojunction can extends from the contacts.

Finally, in an attempt to investigate the selective detection of $H_2$, we also run experiments involving different gases including helium, nitrogen, argon and oxygen. Exposing the samples to He, $N_2$ and Ar we do not observe any change on the device resistance or in the CNP. For a more reactive gas like oxygen, as expected,[22,34] we observe an extra *p*-type doping, but no change in the $R_{odd}$ (see Supp. Mat.). Again, this reinforces that, as the best of our knowledge, the *pn* junction is solely changed under $H_2$ exposure. Fig. 3(a) depicts the change of the device resistance ($\Delta R$) at fixed back gate voltage ($V_G = -1.5\,V$) and concentration (20%) as a function of time for He, $N_2$ and Ar gases under a constant gas flow of 500sccm. This result indicates that the modulation of the device resistance asymmetry discussed here is a selective way of detecting molecular hydrogen. In addition, the response time is highly sensitive to hydrogen concentration, demonstrating faster response time for higher molecules concentrations. The same reversible and asymmetric modulation is seen from 0.1% (the lowest concentration limit of our testing system) up to 50% of $H_2$, as shown in Fig. 3(b). In both figures, we define $\Delta R$ as the change of the device resistance relative to the initial resistance, i.e. the device resistance only with Ar. Also, by comparing the different curves in Fig. 3(b), it is evident that the devices respond faster to the adsorption of hydrogen than to the desorption process. Our results show that graphene devices with invasive contacts effectively can work as a selective $H_2$ sensor without any functionalization, even at room temperature.

In summary, we have investigated the *pn* junction in graphene/*h*-BN devices caused by invasive contacts both experimentally and theoretically. Our analysis shows that the *pn* junction can be modulated in an asymmetric and reversible manner by hydrogen molecules at the metal-graphene interface. Our model points out some intrinsic factors that affect high-performance graphene devices: contact interface length and *pn* junction resistance. In addition, our results are valuable for understanding the metal-graphene interface, demonstrating the necessity of incorporating quantum aspects to the contact resistance. Lastly, it shows that graphene with invasive contacts can be used as a selective detector of molecular hydrogen.

Fig. 1

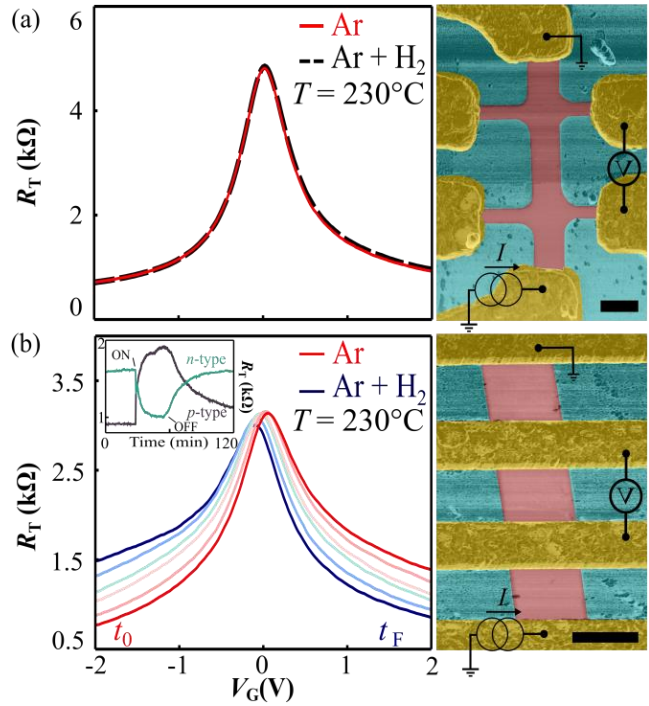

Fig. 1: (a) The total resistance ($R_T$) as a function of back gate voltage ($V_G$) at non-invasive four-terminal graphene device on top of $h$-BN/graphite under argon (Ar) (red curve) and molecular hydrogen ($H_2$) exposure (black dashed curve) at $T=230°C$. The black dashed curve is obtained after 30min of Ar+$H_2$ exposure. (b) $R_T$ vs $V_G$ as a function of the time in the $H_2$ exposure at invasive four-probe device at $T=230°C$. The gradient from red to blue curves indicates the time variation from Ar to Ar+$H_2$ flow; $t_0$ depicts data taken only with Ar before $H_2$ exposure, and $t_F$ is taken after 30min under $H_2$ exposure. The inset shows $R_T$ as a function of time by $H_2$ exposure for electron- ($e$) and hole- ($h$) branch for a fixed back gate voltage, -1.5V and +1.5V, respectively. Turned ON means Ar+$H_2$ (20%) flow, while turned OFF only Ar flow. The right panels are false-color AFM images of devices measured applying current bias. Scale bar, 500nm.

Fig. 2

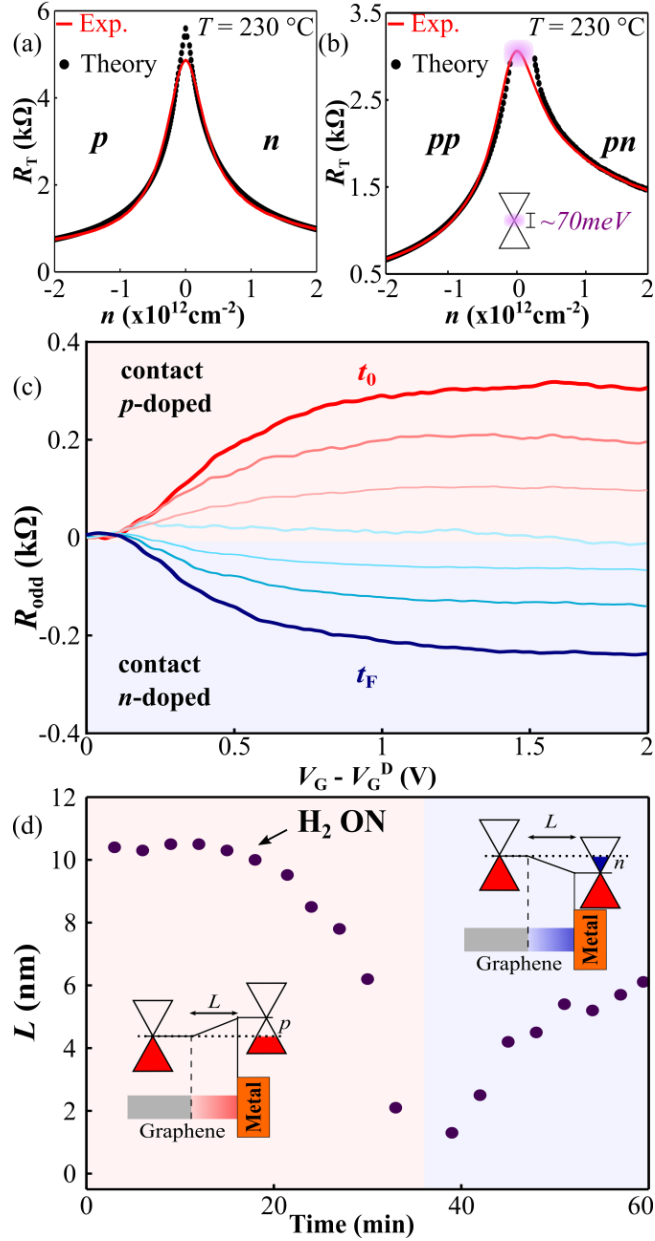

Fig. 2: $R_T$ as a function of charge carrier concentration ($n$) for non-invasive (a) and invasive (b) four-terminal graphene device at $T$=230°C, respectively. The inset in Fig. 2(b) shows the energy fluctuation associated to temperature and device disorder. In both figures the experimental data are shown in red and the fitting are shown in black circles. The asymmetry for invasive probes originates from the quantum contribution of the contact resistance, which thereby forms a *pn* or *pp* junction along the graphene channel during the $V_G$ sweep. Meanwhile, for non-invasive electrodes only a *p* or *n* doping is observed in the graphene channel originated from the back gate applied. (c) The asymmetry between electrons and holes by showing the odd part of the resistance ($R_{odd}$) versus $V_G - V_G^D$; where $V_G^D$ is the point that shows the resistance maxima. The modulation caused by $H_2$ at the interface can be seen clearly by the inversion of the doping induced by the contact, and in the $R_{odd}$. (d) *pn* junction length ($L$) as function of the time during the hydrogen exposure estimated by the fitting using equation 1. Purple circles are the *pn* length estimated by fitting our experimental data. The insets in the Fig. 2(d) show the schematic profile of a *pn* junction along the graphene channel before (left) and after $H_2$ (right) exposure. Gray region indicates the main graphene region, while the light-blue and light-red regions represent the region under the interface potential.

Fig. 3

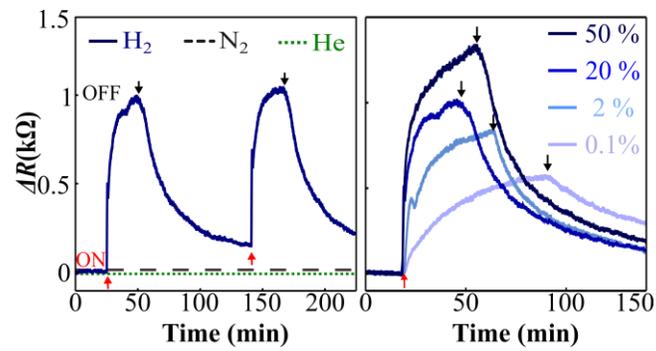

Fig. 3: (a) Changing of device resistance (ΔR) as a function of the time for different gases at a fixed concentration (20%) mixed to argon at $T$=230°C. The solid blue, dashed gray and dotted green curve depicts the mixed with $H_2$, $N_2$ and He, respectively. (b) ΔR vs Time for different $H_2$ concentration. Data was taken at fixed back gate voltage of $V_G$ = -1.5V with graphene main channel *p*-type doped. However, the same time dependence is observed with graphene *n*-type doped.

ACKNOWLEDGMENTS

This work was supported by CAPES, Fapemig, CNPq, Rede de Nano-Instrumentação and INCT/Nanomateriais de Carbono. The authors are thankful to Centro Brasileiro de Pesquisas Físicas (CBPF) and Centro de Componentes Semicondutores (CCS) for providing an e-beam lithography system, and Lab Nano at UFMG for allowing the use of atomic force microscopy.

REFERENCES

[1] A.K. Geim and K.S. Novoselov, Nat. Mater. **6**, 183 (2007).
[2] F. Schwierz, Nat. Nanotechnol. **5**, 487 (2010).
[3] F. Xia, V. Perebeinos, Y. Lin, Y. Wu, and P. Avouris, Nat. Nanotechnol. **6**, 179 (2011).
[4] S.M. Song and B.J. Cho, Carbon Lett. **14**, 162 (2013).
[5] B. Huard, N. Stander, J. a. Sulpizio, and D. Goldhaber-Gordon, Phys. Rev. B **78**, 121402(R) (2008).
[6] H. Xu, S. Wang, Z. Zhang, Z. Wang, H. Xu, and L.-M. Peng, Appl. Phys. Lett. **100**, 103501 (2012).
[7] K. Nagashio, T. Nishimura, K. Kita, and A. Toriumi, in *Int. Electron Devices Meet.* (2009), pp. 565–568.
[8] K. Nagashio, T. Nishimura, K. Kita, and A. Toriumi, Appl. Phys. Lett. **97**, 143514 (2010).
[9] T. Low, S. Hong, J. Appenzeller, S. Datta, and M.S. Lundstrom, IEEE Trans. Electron Devices **56**, 1292 (2009).
[10] E.J.H. Lee, K. Balasubramanian, R.T. Weitz, M. Burghard, and K. Kern, Nat. Nanotechnol. **3**, 486 (2008).
[11] Y. Wu, V. Perebeinos, Y.M. Lin, T. Low, F. Xia, and P. Avouris, Nano Lett. **12**, 1417 (2012).
[12] P. a. Khomyakov, G. Giovannetti, P.C. Rusu, G. Brocks, J. Van Den Brink, and P.J. Kelly, Phys. Rev. B **79**, 1 (2009).
[13] G. Giovannetti, P. a. Khomyakov, G. Brocks, V.M. Karpan, J. Van Den Brink, and P.J. Kelly, Phys. Rev. Lett. **101**, 4 (2008).
[14] V.E. Calado, S. Goswami, G. Nanda, M. Diez, a. R. Akhmerov, K. Watanabe, T. Taniguchi, T.M. Klapwijk, and L.M.K. Vandersypen, Nat. Nanotechnol. **10**, 761 (2015).
[15] R. Nouchi and K. Tanigaki, Appl. Phys. Lett. **106**, 083107 (2015).
[16] S. Wang, D. Mao, Z. Jin, S. Peng, D. Zhang, J. Shi, and X. Wang, Nanotechnology **26**, 405706 (2015).
[17] Z. Zhang, Q. Xue, Y. Du, C. Ling, and W. Xing, J. Mater. Chem. A **2**, 15931 (2014).
[18] P. Blake, R. Yang, S. V. Morozov, F. Schedin, L.A. Ponomarenko, A.A. Zhukov, R.R. Nair, I. V. Grigorieva, K.S. Novoselov, and A.K. Geim, Solid State Commun. **149**, 1068 (2009).
[19] T. Mueller, F. Xia, M. Freitag, J. Tsang, and P. Avouris, Phys. Rev. B **79**, 1 (2009).
[20] I. Barcelos, A. Cadore, L. Campos, A. Malachias, K. Watanabe, T. Taniguchi, F. Barbosa Maia, R. de O. Freitas, and C. Deneke, Nanoscale **7**, 11620 (2015).
[21] A.R. Cadore, E. Mania, K. Watanabe, T. Taniguchi, R.G. Lacerda, and L.C. Campos, arXiv:1603.04872 (2016).
[22] I. Silvestre, E.A. De Morais, A.O. Melo, L.C. Campos, A.M.B. Goncalves, A.R. Cadore, A.S. Ferlauto, H. Chacham, M.S.C. Mazzoni, and R.G. Lacerda, ACS Nano **7**, 6597 (2013).
[23] D. Henwood and J.D. Carey, Phys. Rev. B **75**, 245413 (2007).
[24] B.H. Kim, S.J. Hong, S.J. Baek, H.Y. Jeong, N. Park, M. Lee, S.W. Lee, M. Park, S.W. Chu, H.S. Shin, J. Lim, J.C. Lee, Y. Jun, and Y.W. Park, Sci. Rep. **2**, 1 (2012).
[25] C.R. Dean, A.F. Young, I. Meric, C. Lee, L. Wang, S. Sorgenfrei, K. Watanabe, T. Taniguchi, P. Kim, K.L. Shepard, and J. Hone, Nat. Nanotechnol. **5**, 722 (2010).
[26] E.H. Hwang, S. Adam, and S. Das Sarma, Phys. Rev. Lett. **98**, 186806 (2007).
[27] P.E. Allain and J.N. Fuchs, Eur. Phys. J. B **83**, 301 (2011).
[28] A.F. Young and P. Kim, Nat. Phys. **5**, 222 (2009).
[29] V. V. Cheianov and V.I. Fal'ko, Phys. Rev. B **74**, 1 (2006).
[30] P.A. Khomyakov, A.A. Starikov, G. Brocks, and P.J. Kelly, Phys. Rev. B **82**, 1 (2010).
[31] T.W. Kohl, Claus-Dieter, *Gas Sensing Fundamentals*, Springer S (Springer Berlin Heidelberg, Berlin, Heidelberg, 2014).
[32] L. Barrio, P. Liu, J. a. Rodríguez, J.M. Campos-Martín, and J.L.G. Fierro, J. Chem. Phys. **125**, 12 (2006).
[33] F. Šrobár and O. Procházková, in *8th Int. Conf. Adv. Semicond. Devices Microsystems* (2010), pp. 275–278.
[34] S. Ryu, L. Liu, S. Berciaud, Y.-J. Yu, H. Liu, P. Kim, G.W. Flynn, and L.E. Brus, Nano Lett. **10**, 4944 (2010).

# Supplementary Material
## Metal-Graphene Heterojunction Modulation via $H_2$ interaction

A. R. Cadore,[1,a)] E. Mania,[1] E.A. Morais,[2] K. Watanabe,[3] T. Taniguchi,[3] R. G. Lacerda,[1] and L. C. Campos[1,a)]

[1]*Departamento de Física, Universidade Federal de Minas Gerais, Belo Horizonte, 30123-970, Brazil*

[2]*Universidade Federal de Itajubá, Campus Avançado de Itabira, Itabira, 35903-087, Brazil*

[3]*Advanced Materials Laboratory, National Institute for Materials Science, 1-1Namiki, 305-0044, Japan*

[a)] *Electronic mail: alissoncadore@gmail.com; lccampos@fisica.ufmg.br*

### I – Device resistance as a function gate voltage at several temperatures

Using an invasive contact configuration, we performed measurements of resistance as a function of several temperatures ranging from room temperature up to 230°C. Figures S1(a-c) show the experimental data taken at the following conditions: constant gas flow of 500sccm of molecular hydrogen (concentration of 20%) at three different temperatures (room temperature (25°C), 150°C and 230°C respectively).

Initially, the sample is under 500sccm of ultra-pure Ar (Fig. S1(a)), and shows an asymmetrical resistance as a function of the gate voltage, with a larger resistance for *n*-type doped graphene ($V_G > 0$ V). With addition of molecular hydrogen, the device resistance at the *n*-type branch decreases while the device resistance at the *p*-type branch ($V_G < 0$ V) increases. This behavior happens for all the temperatures but is enhanced at higher temperatures, showing that the tuning of the contact resistance induced by hydrogen molecules is thermo activated. Also, the time response is faster at high temperatures. At room temperature, it takes 3 h until the device resistance saturates and stops to vary under Ar+$H_2$ atmosphere. Moreover, the hydrogen desorption process takes an even longer period – approximately it takes 12h for a complete return of the device electrical properties to its initial conditions. On the other hand, at 150°C it takes about 60min after hydrogen exposure for the device to reach the saturation point. The hydrogen desorption process is also longer, but now it takes about 6h to restore to initial conditions. At 230°C, the device resistance reaches its saturation point about 30min after hydrogen exposure, taking about 3h to restore to it is initial conditions. It is important to point out that the adsorption and desorption times also depend on the hydrogen concentration as shown in the main text.

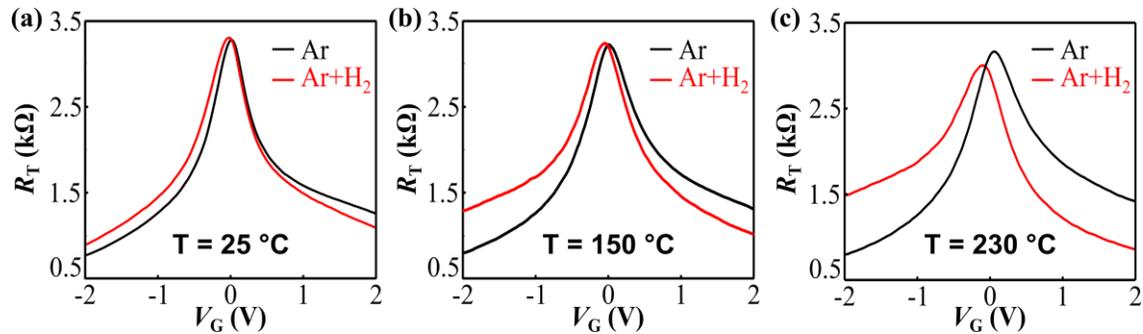

Fig. S1: Total resistance ($R_T$) of graphene devices with invasive contacts as a function of back gate ($V_G$) at: (a) Room temperature; (b) 150°C and; (c) 230°C. The black curves indicate the measurements under argon (Ar) and the red curves show the mixed flow of Ar+$H_2$.

**II – Modeling the device resistance for invasive contacts**

In the case of devices containing invasive contacts, there are significant contributions of the electrodes, i.e., *pn* junction to the device resistance. Thus, to model the device resistance, we consider electrostatic effects at the interface between graphene and contacts. Nearby the electrodes, the Fermi energy monotonically changes from the main channel up to the contact (see Fig. S2(a)). When the imbalance between the electrostatic potentials at the interface leads to a formation of a heterojunction, an additional resistance has to be considered. It is important to mention that a heterojunction in graphene behaves slightly differently from heterojunction in conventional semiconductors. For instance, due to conservation of pseudo-spin, transverse chiral charge carriers in graphene get completely transmitted through the barrier – a phenomena known as Klein tunneling.[1,2] However, any transverse carriers are allowed to cross the barrier with a smaller probability depending on their incident angle. At this point, two effects can be considered to determine the transmission probability of the carrier: pseudo-spin mismatch and an approach that considers the tunneling of the charge carriers through the barrier. In our naive approximation, we will ignore conservation of pseudo-spin, and we calculate the transmission probability of oblique carries considering the tunneling over the barrier as it has been described by Cheianov et al.[1–3] Even though pseudo-spin mismatch also contributes to the increase of the device resistance due to the heterojunction.

To model the resistance of a graphene device one has to consider two contributions: the resistance of the main graphene channel and the contact resistance (illustrated in the bottom of the Fig. S2(b)). The main channel, as discussed above, is well described by diffusive Boltzmann theory, while the contact resistance has contributions of the classical Ohm's law as well as a charge transmission through a heterojunction at the graphene-contact interface ($R_{pn}$). In our simple approximation, we consider a symmetrical electrostatic profile at the metal-graphene interface, and we ignore effects due to diffusive contribution through to the metal-graphene interface. Also, for our four-probe terminal devices we do not consider the contribution of the classical contact resistance.[4]

$$R_T = R_{pn} + R[L_o - L] \quad (1)$$

Here $R_T$ is the total resistance of the device, $R[L_o\text{-}L]$ is the main channel resistance calculated by Boltzmann diffusive model, $L_o$ is the length of the graphene main channel, $L$ is the length of the *pn* junction, and $R_{pn}$ is the resistance ascribed for the *pn* junction. The resistance of the interface $R_{pn}$ is calculated via the Landauer formula, where $e^2/h$ is the quantum of conductance and the sum incorporates the transmission probability of each propagating mode across the metal-graphene interface.[1]

$$R_{pn} = \left[\left(\frac{4e^2}{h}\right)\Sigma_{k_y} T(k_y)\right]^{-1} \quad (2)$$

By using Wentzel Krames Brillouin approximation, the transmission probability is given by $T(k_y) = e^{-\pi L k_y^2/k_F}$, where $k_F = \sqrt{\pi n}$ is the Fermi momentum and $k_y = k_F W/2\pi$ is the transverse momentum related to graphene channel width ($W$) and charge carrier density ($n$).[1–3]

Fig. S2(b) shows the resistance across the graphene measured in atmosphere of argon for invasive contacts. With black circles, we show the fitting performed using our model described by equation 1. The good agreement between experiment and theory, as discussed in the main text, reinforces the existence of the *pn* junction at devices with invasive probes.

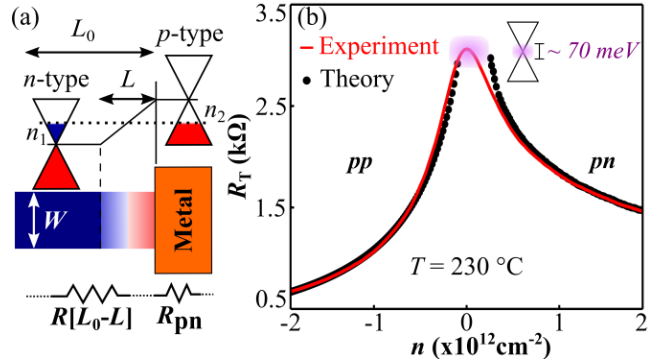

Fig. S2: (a) Schematic view of a *pn* junction along the graphene channel. Blue region indicates the main *n*-type graphene region, while the light-blue and light-red regions represent the region under the interface potential. $L_0$, $L$ and $W$ are the full channel, doped channel length and channel width, respectively. The top part shows a band structure across a smooth *pn* junction. $n_1$ ($n_2$) indicates the doping at the main channel (interface) with its *n* (*p*)-type doping induced by the back gate applied (contact doping). The bottom part represents the resistances ascribed by equation 1. (b) $R_T$ x $n$ for invasive four-probe device at $T$=230°C. The inset shows the energy fluctuation associated to temperature and device disorder. In Fig. (b) the experimental data is shown in red and the fitting is shown in black circles. The asymmetry for invasive probes originates from the quantum contribution of the contact resistance, which thereby forms a *pn* or *pp* junction along the graphene channel during the $V_G$ sweep.

**III – Limitations of the Boltzmann theory at high temperatures**

The device resistance, ignoring effects due to contact resistances, is well described by the Boltzmann transport diffuse model.[5] However, there are some discrepancies between experimental data and Boltzmann diffusive theory near the charge neutrality point (CNP) due to thermal fluctuations.[6] This can be observed by comparing the fitting using Boltzmann theory in our data taken at -269°C (4K) and 230°C (500K) – see Figure S2. At 4K, thermal fluctuation can be neglected and a good agreement between the Boltzmann theory and our experimental data is observed, even close to the CNP. At higher temperatures, as shown here for data taken at 500K, Boltzmann theory converges to the experimental data except at the CNP. The maximum resistance is strongly affected by thermal fluctuations and diverges from the theory. This effect is very evident in the fittings depicted in the Fig, S2(b), where an acceptable agreement between our theory and the data is only possible away from the CNP.

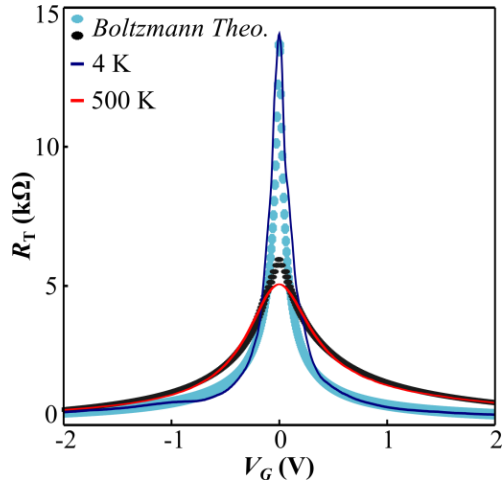

Fig. S2: Total resistance ($R_T$) versus back gate voltage ($V_G$) for measurements taken at 4K and 500K. The circles are the best fits for each temperature.

## IV – Measurements during oxygen exposure

Using an invasive contact configuration, we also performed measurements at 230°C with 10% of oxygen into the chamber. For this procedure, the total gas flow was also 500sccm. Fig. S3(a) shows $R_T$ x $V_G$ for a two probe device. In red, we depict the total resistance under pure Ar atmosphere, and in blue we show data taken under Ar+$O_2$. The charge neutrality point shifts positively after $O_2$ exposure, indicating a p-type doping, and also there are changes in the charge carrier mobilities: hole mobility increases and the electron mobility decreases, as discussed elsewhere.[7,8] However, as shown in Fig. S3(b), the odd resistance does not change after the oxygen exposure. Therefore, it indicates that $O_2$ does not affect the *pn* junction underneath the contacts and only interacts with the graphene main channel.

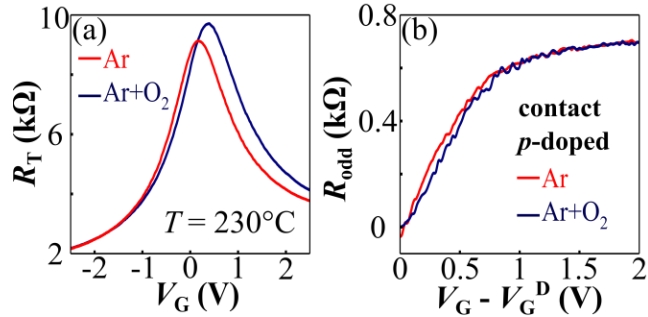

Fig. S3: (a) $R_T$ x $V_G$ at 230°C. The red curve indicates the measurements under Ar and the red curve show the mixed flow of Ar+$O_2$. (b) The asymmetry between electrons and holes by showing the odd part of the resistance ($R_{odd}$) versus $V_G$ - $V_G^D$; where $V_G^D$ is the point that shows the resistance maxima. From this figure, one can easily observe no modification caused by $O_2$ at the interface, keeping the same *p*-type doping induced by the gold contacts.

REFERENCES

[1] P.E. Allain and J.N. Fuchs, Eur. Phys. J. B **83**, 301 (2011).
[2] A.F. Young and P. Kim, Nat. Phys. **5**, 222 (2009).
[3] V. V. Cheianov and V.I. Fal'ko, Phys. Rev. B **74**, 1 (2006).
[4] S.M. Song and B.J. Cho, Carbon Lett. **14**, 162 (2013).
[5] C.R. Dean, A.F. Young, I. Meric, C. Lee, L. Wang, S. Sorgenfrei, K. Watanabe, T. Taniguchi, P. Kim, K.L. Shepard, and J. Hone, Nat. Nanotechnol. **5**, 722 (2010).
[6] J. Schiefele, F. Sols, and F. Guinea, Phys. Rev. B **85**, 195420 (2012).
[7] S. Ryu, L. Liu, S. Berciaud, Y.-J. Yu, H. Liu, P. Kim, G.W. Flynn, and L.E. Brus, Nano Lett. **10**, 4944 (2010).
[8] I. Silvestre, E.A. De Morais, A.O. Melo, L.C. Campos, A.M.B. Goncalves, A.R. Cadore, A.S. Ferlauto, H. Chacham, M.S.C. Mazzoni, and R.G. Lacerda, ACS Nano **7**, 6597 (2013).